\documentclass[11pt,preprint]{aastex}
\newcommand{\etal}{{\it et~al.}}

\begin{document}

\title{Palomar Optical Spectrum of Hyperbolic Near-Earth Object A/2017 U1}

\author{Joseph R. Masiero\altaffilmark{1}}

\altaffiltext{1}{Jet Propulsion Laboratory/California Institute of Technology, 4800 Oak Grove Dr., MS 183-301, Pasadena, CA 91109, USA, {\it Joseph.Masiero@jpl.nasa.gov}}

\begin{abstract}
We present optical spectroscopy of the recently discovered hyperbolic
near-Earth object A/2017 U1, taken on 25 Oct 2017 at Palomar
Observatory.  Although our data are at a very low signal-to-noise,
they indicate a very red surface at optical wavelengths without
significant absorption features.

\end{abstract}

\section{Introduction}

On 25 Oct 2017, the Minor Planet Center (MPC) published a Minor Planet
Electronic Circular (MPEC) announcing the discovery of object C/2017
U1\citep{MPEC1}, which was subsequently redesignated as A/2017 U1
after no cometary activity was seen in deep imaging stacks
\citep{MPEC2}.  A/2017 U1 has a strongly hyperbolic orbit
(eccentricity$=1.191\pm0.007$), with a $v_\infty$ with respect to the Sun
of $\sim25~$km/s, indicating that it likely originated from beyond our
Solar System.  This is the first known minor planet energetically
unbound from the Sun.

At the time of the announcement, we were observing on the Hale $5~$m
telescope at Palomar mountain using the Double Spectrograph (DBSP)
optical spectrograph \citep{oke82}. We present the spectral
observations of A/2017 U1 we obtained during this run below.

\section{Data}

All of our observations were obtained on 25 Oct 2017 UT, shortly after
A/2017 U1 was announced.  On this date A/2017 U1 was at a phase angle
of $\alpha=19.5^\circ$, a geocentric distance of $\Delta=0.40~$AU, and a
heliocentric distance of $R_\sun=1.36~$AU. At the time of observation,
A/2017 U1 had an apparent magnitude of $V\sim21~$mag, making it a difficult
target even for Hale.  Further, the atmospheric seeing at the time of
observations was poor and variable ($2-3''$) resulting in observations
with no detectable flux above the background level.  Of the $\sim3$
hours of time spent integrating on the object, only one-third resulted
in useful data where flux from the object was visible on the chip at
all.  In total, we combined ten exposures of $300~$seconds on
integration each for our final spectrum.  Useful observations ran from
06:13 to 09:17 (UT) on 25 Oct 2017.

The DBSP instrument has two arms, split with a dichroic beamsplitter
into red and blue components.  We used low resolution gratings in each
arm ($300$ lines/mm in the blue and $158$ lines/mm in the red), and
the D52 dichroic that results in a split at $520~nm$.  The resulting
unbinned spectral resolution is $R_{blue}=491$ and $R_{red}=488$ at
the blaze wavelengths of $399~$nm and $756~$nm, respectively.  No
filter was used for either arm, to maximize flux.  We used a $1.5''$
slit width, and aligned the slit along the parallactic angle to reduce
the effect of atmospheric differential refraction.

We performed bias and flat field calibration, as well as wavelength
calibration using a He-Hg arc lamp and a He-Ne-Ar arc lamp for the
blue and red sides (respectively).  Reflectance spectra were obtained
by observing a local standard star (in this case BD+03 27, a nearby
$V=9.4~$mag G0 star) to correct for local atmospheric extinction, and
then slope-corrected with observations of a Solar analog star (here HD
1368, a $V=8.9~$mag G2V star).

For the object and each standard star, we extract a spectrum from the
chip based on a visual identification of the pixels that the spectrum
falls on.  During observing the targets are placed on the same part of
the slit each time, so this region typically varies by only $\sim10$
pixels from object to object.  We used a $30$-pixel wide
extraction region for the standard stars to encompass as much light as
possible, while we used a $12$-pixel wide region for A/2017
U1 to minimize the contribution of the background.  A sky region was
extracted above and below the spectrum ranging from 10 to 30 pixels
away from the edge of our spectral extraction window in both
directions.  These sky measurements were median-combined and
subtracted from each measured spectrum.  Each spectrum from
individual exposures is then median-combined in wavelength-space to
get the final measured spectrum.  We divide the object spectrum by the
local standard in wavelength space to correct for atmospheric
conditions.  To correct for the spectral type of the local standard,
we divide the local standard by the solar standard, perform a linear
fit, then divide the corrected object spectrum by this linear trend.

We restrict our analysis to wavelengths from $520~$nm to $950~$nm.
Blueward of this range the light is primarily sent to the blue camera,
which results in different systematics and requires more detailed
calibration.  In addition, no flux can be seen by eye at these
wavelengths in the raw image frames.  Redward of our range of
interest, telluric lines from the atmosphere begin to significantly
contaminate our spectra resulting in a large increase in noise.

\begin{figure}[ht]
\begin{center}
\includegraphics[scale=0.7]{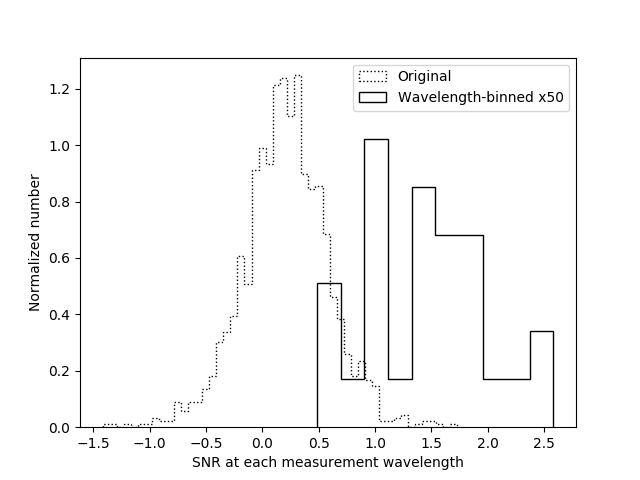}
\protect\caption{Histograms of the signal-to-noise at each wavelength
  element for the full median-combined spectrum (dotted), and the same
  data binned by 50 elements in wavelength space with an
  error-weighted mean (solid).}
\label{fig.snr}
\end{center}
\end{figure}

In Figure~\ref{fig.snr} we show the signal-to-noise (flux over error)
of each wavelength element in our final object reflectance spectrum.
There is a distinct positive trend in this distribution, showing that
we did detect flux from A/2017 U1.  By binning the data in
wavelength-space using an error-weighted mean we can improve the
signal-to-noise of each element, as shown by the solid line in
Fig~\ref{fig.snr}.  

\begin{figure}[ht]
\begin{center}
\includegraphics[scale=0.6]{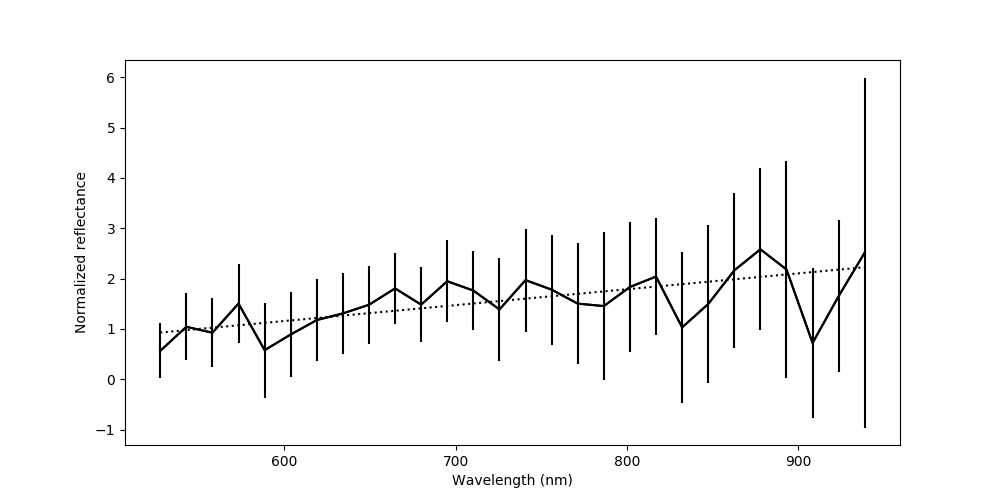}
\protect\caption{Final normalized, binned reflectance spectrum of
  A/2017 U1 (solid line).  Error bars are shown for each binned
  wavelength point, and a best-fit linear trend is shown as a dotted
  line. The fitted slope is $3.0\pm1.5$ normalized reflectance units
  per micron (equivalent to $30\%/1000\AA$.}
\label{fig.spec}
\end{center}
\end{figure}

We show in Figure~\ref{fig.spec} our calibrated, binned spectrum of
A/2017 U1.  We have normalized the spectrum to the measured
reflectance at $550~$nm, and measure a slope following Equation 1 from
\citet{bus02}.  No significant deviation from a linear trend is
observed, and the best-fit line has a slope of
$3.0\pm1.5~$units/$\mu$m (equivalent to $30\% / 1000\AA$).

\section{Results}

Our spectrum of A/2017 U1 shows no significant features, however our
data have a low enough signal-to-noise that this is not a conclusive
determination.  The spectral slope is only measured at $2~\sigma$ from
a line of zero-slope, but is significantly redder than any
of the taxonomies found in the SMASS survey \citep{bus02}, and
consistent with the RR class seen in the TNOs and Centaurs
\citep{merlin17}.

Integrating our measured reflectance spectrum over the SDSS bandpasses
\citep{fukugita96}, we derive colors of $g-r=0.2\pm0.4$,
$r-i=0.3\pm0.3$, and $r-z=0.4\pm0.4$ which are redder than the bulk of
asteroids measured by SDSS \citep{ivezic01} and are consistent with
the Kuiper Belt colors measured in the Col-OSSOS survey
\citep{pike17}.

Due to the low SNR of our observations, we can only make very limited
interpretations of these data.  However within that bound, A/2017 U1
appears to have a red slope at optical wavelengths and no significant
absorption features in this wavelength range.

\section*{Acknowledgments}

I would like to thank the contributors on the Minor Planet Mailing
List and Michele Bannister for bringing this object to my attention.
I also thank Davide Farnocchia for providing critical late-night,
real-time ephemerids for this object which allowed it to be observed,
and Kajsa Peffer and Paul Nied for support during observing.  This
research was carried out at the Jet Propulsion Laboratory, California
Institute of Technology, under a contract with the National
Aeronautics and Space Administration. This research has made use of
data and services provided by the International Astronomical Union's
Minor Planet Center.  This publication makes use of observations from
the Hale Telescope at Palomar Observatory which is owned and operated
by Caltech, and administered by Caltech Optical Observatories.

\end{document}